# Experimental evidence of optical analogue of electromagnetically induced transparency in graphene

Lei Gao, Tao Zhu[*], Min Liu, Wei Huang, Yu Jia Li, and Cong Gao

Key Laboratory of Optoelectronic Technology & Systems (Ministry of Education), Chongqing University, Chongqing 400044, China.
[*]zhutao@cqu.edu.cn

We demonstrate the first experimental observation of coherent population oscillation, an optical analogue of electromagnetically induced transparency, in graphene based on phase sensitive pump-probe system. Degenerate four-wave-mixing between pump and probe modifies the dispersion and absorption of the population oscillation process in graphene, and leads to enhance and depression of modulation instability with asymmetry in frequency. The analytically predicted asymmetrical burning hole fully consists with the experiments.

    Since the first observation in atomic gases [1], the electromagnetically induced transparency (EIT), a quantum interference effect where a sharp transparency window appear within a broad absorption spectrum when excited by a high driven laser field, has shown intriguing potentials in slow light, quantum processing, and optical switching [2-5]. To refrain from the extreme quantum conditions, great efforts have made to investigate the EIT-like behaviors in bulk solid states at room temperature, and numerous systems have been proposed, such as coupled microresonator, stimulated Brillouin scattering, photonic crystal, metallic plasmonic structures, and coherent population oscillation (CPO) [6-10]. CPO is referred to the oscillation of the ground state population with a beat frequency between pump and probe fields in a saturable medium, where a narrow spectral hole occurs in a homogeneously broadened absorption spectrum with a linewidth on the order of the inverse of the relaxation time. This theory is frequently used for explaining the "slow light" in the erbium-doped fiber and the ruby: a sharp spectral change in the transmission curve results in a strong change in the group velocity [3, 10, 11]. For example, the ground state recovery time of 4.45 ms in ruby produces a spectral hole with a full width at half maximum (FWMH) width of 35.8 Hz [3, 12]. However, such a narrow burning hole limits the operating bandwidth and makes it extremely difficult to perform direct pump-probe measurement due to the limited fine wavelength tuning of pump. As argued by Selden, many apparently slow light phenomenon actually do not meet the conditions necessary for CPO-based slow light, i.e. detuning of pump and probe is within the narrow homogeneous linewidth (inverse of the relaxation time). Instead, he explained them with the saturable absorption phenomena, an intensity-dependent and frequency-independent nonlinear process, and full agreement was achieved [13]. Therefore, it is necessary to find more saturable absorber materials with a sufficient small relaxation time (corresponding to a wide burning hole).

    It has been shown that graphene possesses unique linear energy dispersion. When it is excited by laser with photon energy near the Dirac-zone, the electron is excited from valence band to conduction band, then they thermalize and cool down to form a Fermi-Dirac distribution. The relaxation time corresponding to carrier-carrier intraband collisions and phonon emission is about 100 fs, and the relaxation time associated with electron interband relaxation and cooling of hot photons is ~ 3 ps [15]. For graphene, the Pauli blocking prevents further absorption, leading to the saturable absorption effect [14,15]. Besides, the bandgap

structure of graphene can be tuned dynamically by gating voltage or chemical doping, through which the conductivity as well as the transmission can be changed. The field-control capability motivates diverse devices with extraordinary performances, such as phase modulator [16], and plasmonic analogs of EIT configurations based on graphene-related waveguides [9, 17]. However, graphene is treated as an alternative material to metal to excite localized plasmons, rather than the direct material for EIT-like phenomenon. Motivated by Pauling exclusion, all optical modulation was proposed [18,19]. Sometimes it was called the cross absorption modulation (XAM). As illustrated schematically in Supplementary 1, when a strong pump laser illuminates graphene, the formed Fermi-Dirac distribution blocks the interband optical transitions, and prevents absorption of probe with energy lower than that of pump. The modulation depth and insertion loss of graphene decreases when increasing the power of external CWL with pump-probe wavelength shift of 570 nm or 490 nm [18,19]. However, XAM cannot exist when the probe energy is larger than that of the pump, and we find that another effect can occur once the pump-probe frequency detuning is smaller than the inverse of the relaxation time of graphene.

In this Letter, we identify the EIT-like behavior in graphene utilizing a phase sensitive pump-probe system. According to the CPO theory, the FWHM width of the burning hole is $1/2\pi*\tau$, where $\tau$ is the relaxation time of the electrons. Correspondingly, the width of burning hole in graphene is about 53 GHz (~0.5 nm in C band) at room temperature. However, there are two problems with the direct pump-probe measurement. One is the small frequency-independent absorption of graphene (0.023). It is hard to detect a little variation on such a small loss with pump-probe detuning of 0.5 nm, especially it coexists with XAM, an effect that would cover the absorption variation from CPO. The other is that the broad burning-hole makes the group velocity variation even much smaller with the spreading of 53 GHz. In order to solve the problems, we have designed a special phase sensitive system to observe the CPO effect in graphene clearly.

The main setup is schematically shown in Fig. 1, which is used previously for partially mode-locked fiber laser, and its formation process based on vector four-wave-mixing (FWM) among parametric gain lobes has been explained in detail in [20]. A well fit of Lorentzian shape of 2D band in the Raman spectrum implies that the graphene flakes are electronically decoupled, and they maintains Dirac fermions linear dispersion [15]. The modulation depth of the SA is ~ 3%. We couple an external CWL with flexible tunability of polarization, wavelength, and power, into the cavity with a counterclockwise direction. As the external CWL circulates in the cavity, it passes through the SA many times. Besides, we conduct a conditional control experiment, where the CWL is coupled to the main cavity between SA and DCF, but with a clock wise direction. In this case, the CWL passes through SA only once, then it is blocked by ISO. Similar results can be obtained, but requires a much larger power (> 100 mW). The line width of the external CWL is 0.05 nm, and its power is much smaller than that of the system, hence, its influence on the gain property of the EDF can be ignored. It can be seen that the system output results from the variation of phase matching condition due to the nonlinear phase shift of SA under the illumination of external CWL.

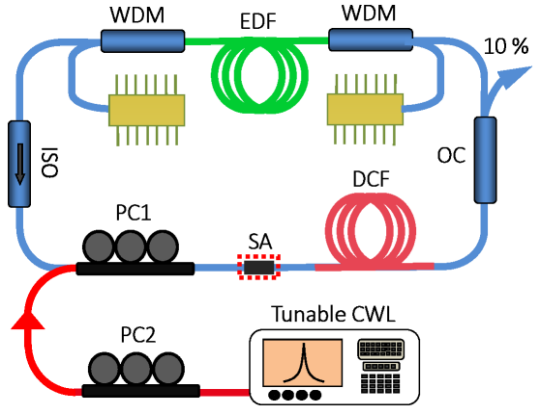

Figure 1. Setup of the pump-probe system. EDF, erbium-doped fiber; WDM, wavelength division multiplexer; ISO, polarization independent optical isolator; PC, polarization controller; DCF, dispersion compensation fiber; OC, optical coupler; SA, saturable absorber; CWL, continuous wave laser.

With proper cavity detuning, this system may generate primary gain lobes alone, namely, the excitation of parametric instability (PI). Physically, it is a degenerate four-wave-mixing process (FWM), and the quasi-phase-matching condition can be described as $2\gamma P + \bar{\beta}_2 \Delta\omega_k^2 = 2k\pi/Z$, where, γ, P, β$_2$, Z are nonlinear coefficient, peak power, average group velocity dispersion, and dispersion period, respectively [21,22]. Δω is the frequency shift from the center, and $k$ presents the sidebands order. PI occurs when the phase accumulated in each round trip equals to 2$k$π. The sidebands can be depressed with the cavity being off resonance through changing the phase by biasing PC, or modulating graphene with external CWL based on CPO effect.

We set two 976 lasers to 400 mW, and the power of the external CWL as zero, and we obtain PI gain sidebands when rotating PC properly. The optical spectrum of the PI shown in Fig. 2 is denoted as original, from an operating state similar to the PC state 2 in Ref. 20. The two sidebands are generated by degenerate FWM. Here, we find that the two sidebands are relatively weak with respect to the center wavelength, and the state of polarization (SOP) of the output is fixed as a closely clustered spot in the Poincaré sphere.

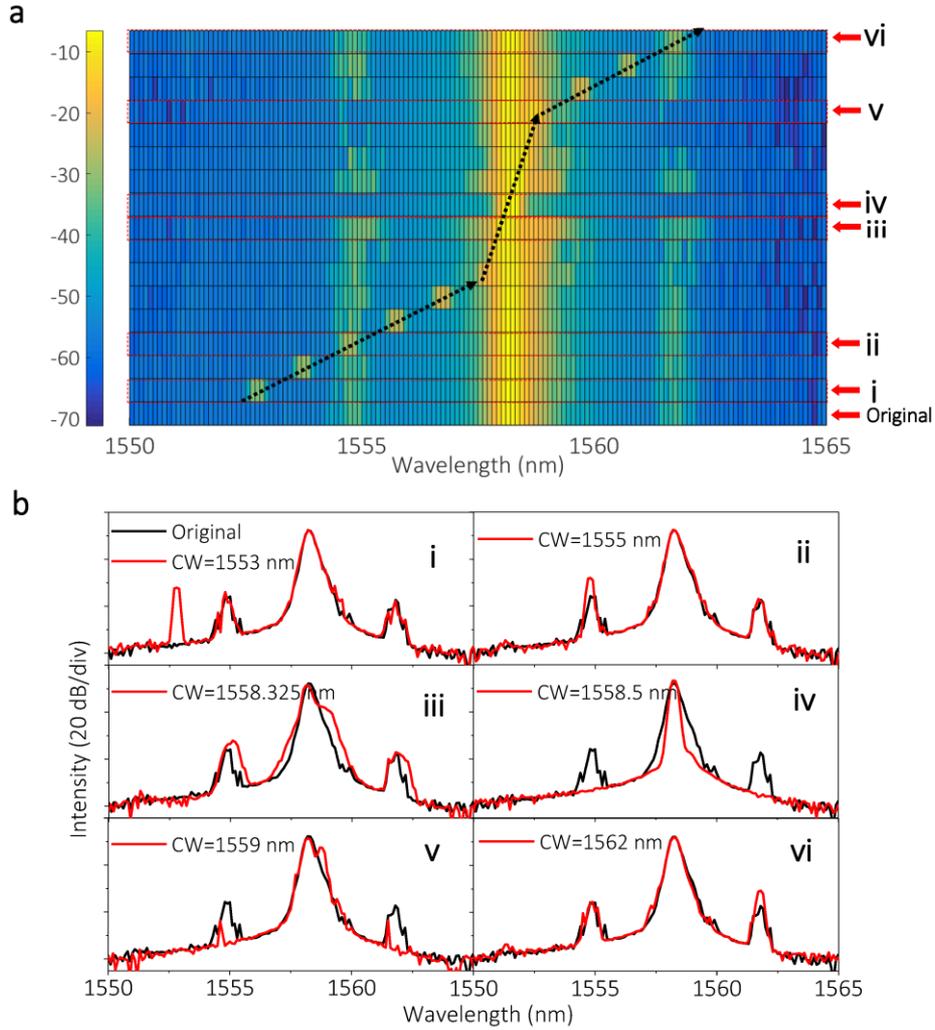

Figure 2. (a) Total optical spectra of the PI under different CWL wavelengths, the intensities are shown with dB scale. The dotted line present the center wavelength detuning of the CWL from blue side to the red side with two step sizes (Far from center region: 1 nm/step; center region: 0.175 nm/step). (b) Spliced optical spectra from (a) with respect to the original spectrum.

The wavelength of the external laser with a power of 1 mW is fixed to 1558.5 nm, which is the center wavelength of PI. The SOP of the CWL is tuned to cover the whole space of the Poincaré sphere, and significant changes occur only when the SOP of the external laser is located on the same spot of the PI in the Poincaré sphere. For CWL at this wavelength, depression of sidebands can only be observed for power higher than 0.5 mW.

The influence of the lasing wavelength of the external laser on PI is investigated with a power of 1 mW, and the SOP is overlapped with that of PI. The optical spectra of the PI under different CWL wavelengths are presented in Fig. 2. When the wavelength of external laser is scanning from 1553 nm to 1562 nm, significant changes can be observed only near the center wavelength region. It should be noticed that the enhancement of PI occurs only on the blue side of the center wavelength (III in Fig 2. (b)) within a wavelength detuning of ~0.175 nm, while great depression of PI appears on the center wavelength (IV in Fig. 2(b)). This depression effect diminishes gradually when the external laser is tuned far away from the center wavelength in the red side with a wavelength detuning of ~0.5 nm.

Considering the relatively narrow detuning frequency region (0.675 nm), the contribution from gain competition and saturable absorption effect from EDF can be neglected reasonably. Besides, the XAM effect in graphene can also be excluded as it is free from the pump-probe detuning range as long as the pump frequency is smaller than that of the probe. In fact, we have conducted a direct pump-probe measurement with detecting the transmission variation under various circumstances, which are presented in the Supplementary 2. However, the obtained data in the Supplementary 2 cannot support firmly the existence of EIT-like effect. The main reason is that the direct pump-probe scheme is phase insensitive, and the small transmission variation may be generated by other effects. Besides, we also a similar pump-probe experiment when the main laser cavity operating with a partially mode-locked laser (PML) state, where cascaded FWM occurs among PI sidebands (PI). The details are in the Supplementary 3, and similar results are obtained. Specifically, for this lasing state with broadband optical spectrum, the SOP of the laser is randomly locating in a pool of the Poincaré sphere, therefore, the suppression of PML require a much SOP tolerance than that of pure PI sidebands. That's to say, depression of PML occurs when the SOP of CWL locates in the SOP pool of PML. We find that the threshold power of CWL is here is also much larger than that of pure PI sidebands. For the schemes in the main manuscript, the phase matching condition of PI is not sensitive to linear loss or gain, but to the phase variation, and the results in Fig. 2 can be interpreted based on the CPO with degenerated FWM.

Similar to the semiconductor quantum well [23], when graphene is excited by a strong pump laser, and a weak probe laser with a detuning smaller than the decay rate of transition, population oscillation of ground state can occur due to the interference of the two beams. As shown in Fig. 3 (a), the oscillation, with a period equaling to the beat frequency, modulates the transmission, and results in a reduction of the absorption of the probe beam. When graphene is excited around C band, it can be treated as a three-level system. As shown in Fig. 3 (b), for few layer graphene, the ground state |1>, is Dirac zero point, and the stable state |3> rapidly transfer into metastable |2> with a relaxation time of 100 fs (dephasing time $T_1$) due to carrier-carrier intraband collisions and phonon emission. The decay time $T_2$, corresponding to state |2> to state|1>, is 3 ps due to electron interband relaxation and cooling of hot photons. As $T_1$ is much larger than $T_2$, this system can be degenerated to a two-level system. Using rate equation analysis, Schweinsberg et al modelled the CPO in EDF, and the absorption, and the phase shift experienced by probe light exhibited symmetry with respect to frequency [10]. This model fails to explain our experiment as the contribution of degenerated FWM to the CPO effect is neglected in rate equation analysis.

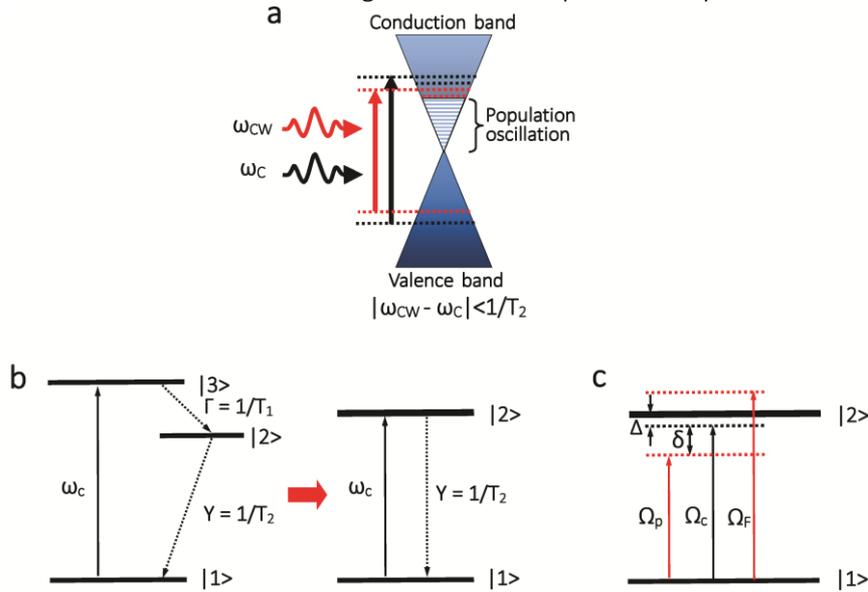

Figure 3 (a) Schematic illustration of CPO effect in graphene with frequency detuning smaller than the inverse of the relaxation time $T_2$. (b) Three-level system of graphene when excited by a laser, which can be simplied to a two-level system. (c) Schematic diagram for a two-level system consisting of a strong center field and two symmetrically tuned weak fields, with Rabi frequency of $\Omega_i$ (i=p, c, F). The pump laser is labeled as "p", and the other is labeled as "F" corresponding to FWM field.

To fully describe the CPO process accompanied by two degenerated FWMs, a density matrix formalism is needed. Fig. 3(c) presents a homogeneously broaden two-level system consisting a center laser, and two weak lasers with symmetrical detuning. The center laser is the giant CW in the system and the pump laser corresponds to the external CWL. The definition of pump laser and FWM laser is arbitrary as the wavelength of external CWL is tuned from the blue side to red one. In the experiment, the frequency mismatching of the center laser and state |2>, Δ, can be neglected, namely Δ=0. δ is the detuning of the two weak laser and center laser. For near-resonat coupling (just as in this experiment), and under electic-dipole and rotating-wave approximation, the analytical steady-state expression of asymmetric density matrix element can be written as [24]:

$$\sigma_{21}^{(+)} = -\frac{iW^{(0)}e^{i\theta_p}}{\Gamma}[(1-\frac{I_c}{D})|\Omega_p| - \frac{I_c|\Omega_F|e^{-i\Phi}}{D}]$$

$$\sigma_{21}^{(-)} = -\frac{iW^{(0)}e^{i\theta_F}}{\Gamma}[(1-\frac{I_c}{D^*})|\Omega_F| - \frac{I_c|\Omega_p|e^{-i\Phi}}{D^*}]$$

(1)

Where, Γ = 1/$T_1$ and Y = 1/$T_2$ are the dephasing rate and decay rate of graphene. Φ = 2$\theta_c$-$\theta_p$ –$\theta_F$, is the relative phase difference of the three fields, D= iδ/Y+1+$I_c$, and population inversion $W^{(0)}$=-1/(1+ $I_c$) with $I_c = 4|\Omega_c|^2/\Upsilon\Gamma$. Two terms are included in Eq. (1): the first term represents the pure CPO in the absence of other two fields, and the other contributes from FWM. The probe field is degenerated with the FWHM field, that is, the Rabi frequencies of pump field and FWHM field are the same, leading that , the pure CPO and FWM contribute equally to the whole effect. The phase-dependent response leads to assymetry of the CPO, and also provides controlability of the CPO process. Generally, we can assume $\theta_p$=$\theta_F$=0, the real part and image part of the susceptibility, corresponding to dispersion and absorption properties of the system, are given by

$$\text{Re}[\chi_{21}^{(+)}(\omega)] = \frac{N|\mu_{21}|^2 W^{(0)}}{2\varepsilon_0\hbar\Gamma}[\frac{I_c(1+R\cos(\theta))\delta/\Upsilon - (I_c+1)R\sin(\theta)}{|D|^2}]$$

$$\text{Im}[\chi_{21}^{(+)}(\omega)] = \frac{N|\mu_{21}|^2 W^{(0)}}{2\varepsilon_0\hbar\Gamma}[1 - \frac{I_c(I_c+1)(1+R\cos(\theta)) + R\sin(\theta)I_c\delta/\Upsilon}{|D|^2}]$$

(2)

Where, R = $\Omega_F$/$\Omega_p$, is the ratio of the Rabi frequency, and $\mu_{21}$ and N are the dipole and atomic density, Γ= 30Y, $I_c$=2.86, $\Omega_c$=0.0155Γ, and $N|\mu_{21}|^2/2\varepsilon_0\hbar = \Gamma$. We plot the imaginary part and real part of $\chi_{21}$ as a function of δ and θ in Fig. 4. It can be observed that the variation of the phase difference, θ, leads to the asymmetry of the burning hole of the CPO. When θ=0, the depth of burning hole reach to maximum due to the construcutre interference of CPO and FWM, while the burn hole disappears when θ=π for the destrucutvie interference [24-27]. When 0<θ<π, for example θ=π/2, both the dispersion and absorption lines show great assymetry. Specifically, the absorption of the system can be manipulated either positive or negative by this phase detuning. Those results are fully agree with our experimental results that both enhancement and depression of PI are observed with asymmetry in frequency, as the intense variation of dispersion changes the phase mathcing condition of PI. However, the enhancement and depression of PI disappers when the external CWL has different polarization from that of the laser system, as the external CWL modifies CPO only when they are inphase, in a case significant FWM occurs between the external CWL and the center laser.

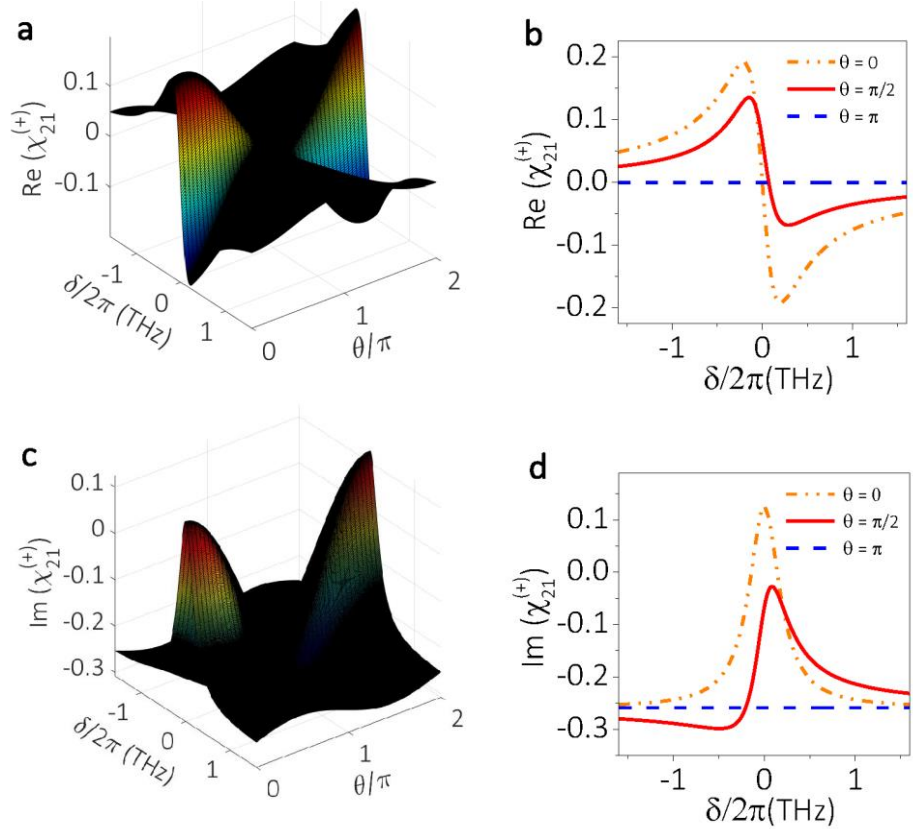

Figure 4 Susceptibility as a function of different detuning and phase differences. (a) and (b) are real part (dispersion) of $\chi_{21}$, while (c) and (d) are the image part (absorption) of $\chi_{21}$.

In summary, we have proposed a pump-probe system to investigate CPO, an EIT-like effect, in graphene. The phase sensitivity of PI is utilized to remove the contributions from XAM. We have found that the degenerate FWM of external CWL and center laser modifies the conventional CPO, and an asymmetrical burning hole is obtained due to their phase difference. The discovery of EIT-like effect in graphene may inspire further investigation of photonic devices for controllable slow light, quantum switching, and photon storage, and optical process.

Acknowledgments

This work was supported by Natural Science Foundation of China (No. 61405020, 61475029, and 61377066) and the Science Fund for Distinguished Young Scholars of Chongqing (No.